\documentstyle[emulateapj,psfig]{article}
\def\lsim{\lower.5ex\hbox{$\; \buildrel < \over \sim \;$}}
\def\gsim{\lower.5ex\hbox{$\; \buildrel > \over \sim \;$}}
\lefthead{}
\righthead{Pseudo-Potential for Accretion Disk around Kerr Black Holes}

\def\lsim{\lower.5ex\hbox{$\; \buildrel < \over \sim \;$}}
\def\gsim{\lower.5ex\hbox{$\; \buildrel > \over \sim \;$}}

\begin{document}

\title{Description of Pseudo-Newtonian Potential for the Relativistic 
Accretion Disk around Kerr Black Holes}
 
\author{Banibrata Mukhopadhyay}

\affil{Inter-University Centre for Astronomy and Astrophysics,
Post Bag 4, Ganeshkhind, Pune-411007, India}

\begin{abstract}

We present a pseudo-Newtonian potential for accretion disk modeling
around the rotating black holes. This potential can describe the general
relativistic effects on accretion disk. As the inclusion of rotation
in a proper way is very important at an inner edge of disk the potential is
derived from the Kerr metric. This potential can 
reproduce all the essential properties of general relativity
within $10\%$ error even for rapidly rotating black holes.

\end{abstract}


\section{Introduction}

Most of the theoretical studies of general relativity in astronomy are approached by a Newtonian or
pseudo-Newtonian method. To avoid the complexity of full general relativistic
equations it is simpler to use non-relativistic equations but with
the inclusion of corresponding (pseudo) potential which can reproduce
some relativistic effects according to the geometry of space-time. Using this potential one can get the approximate 
solutions of the hydrodynamical equations. Shakura \& Sunyaev (1973) initiated the modeling 
of accretion disk around black holes using Newtonian gravitational potential as
\begin{eqnarray}
V_0=-\frac{1}{x},
\label{ss} 
\end{eqnarray}
where, $x=r/r_g$, $r$ is the radial coordinate of the disk and $r_g=GM/c^2$.
However, this potential can not reproduce the properties of the inner region 
of a disk where relativistic effects become important. 
Later on, Paczy\'nski \& Wiita (1980) proposed a pseudo-Newtonian potential
which can reproduce approximately the properties of the inner disk at close to the equatorial plane
around non-rotating black holes without using relativistic fluid equations as
\begin{eqnarray}
V_1=-\frac{1}{(x-2)}.
\label{pw} 
\end{eqnarray}
The beauty of their potential is that, it can give the right positions of the marginally stable ($x_s$)
and marginally bound ($x_b$) orbits in a Schwarzschild metric. It also
reproduces the total mechanical energy per unit mass at the last stable circular
orbit ($E_s$) and the total energy dissipation at a given radius ($\eta$)
in good agreement with that of Schwarzschild geometry (Artemova et al. 1996). 
The error in both cases is less than $10\%$. Nowak \& Wagoner 
(1991) proposed another potential for accretion disk around non-rotating black holes as
\begin{eqnarray}
V_2=-\frac{1}{x}\left[1-\frac{3}{x}+\frac{12}{x^2}\right],
\label{wog} 
\end{eqnarray}
which can reproduce the values of $x_s$ and angular velocity ($\Omega$)
at that radius in Schwarzschild geometry. So far this choice of the potential gives
the best approximate radial epicyclic frequency. Artemova et al. (1996) proposed
two {\it correct potentials} for describing the accretion disk around rotating black holes.
The form of one of their potentials is given as
\begin{eqnarray}
\frac{dV_3}{dx}=-\frac{1}{x^{2-\beta}(x-x_1)^\beta},
\label{art} 
\end{eqnarray}
$x_1$ is the black hole horizon and $\beta$ is a constant for a particular
specific angular momentum of the black hole, $a$ (for exact expression see Artemova et al. (1996)).
They showed that their potentials can reproduce the value of $x_s$ exactly as that for Kerr geometry
and reproduce the values of $\eta$ at different radii in good agreement with that
of general relativistic results. After that, several authors (e.g. Artemova et al. (1996),
Lovas (1998), Semer\'ak \& Karas (1999)) have analyzed the efficiency of 
different pseudo-potentials prescribed for accretion disks around black holes.
In the context of accretion-disk-corona,
which is infalling towards the black hole, Miwa et al. (1998) chose the pseudo-potential
$V_3$ (Eqn. (\ref{art})) and discussed about radiation flux, velocity of infalling corona upto
very close to the black hole.

However, while Eqn. (\ref{art}) is the analytical form for force, other pseudo-potentials for non-rotating 
black holes like Eqns. (\ref{pw}) and (\ref{wog}) have a simple analytical form of potential. 
For the study of parameter space, e.g. sonic point analysis etc. it is useful to have
an analytical expression for the potential which should asymptotically vary as $-1/x$ and
reduce to Paczy\'nsky-Wiita form (Eqn. (\ref{pw})) for zero rotation. Apart from that, 
equation (15) of Artemova et al. (1996) is only valid for co-rotating (positive values of Kerr parameter, $a$) disks.
If one takes negative values of $a$ to represent counter-rotation and using that equation
(15) calculates $E_s$  and $x_b$, the error may be upto $50\%$ and $500\%$ respectively. Similarly, for negative 
$a$ it can not give the correct value of $x_s$. The general
expression of equation (15) of Artemova et al. (1996) should read
$r_{\rm in}=3+Z_2\mp[(3-Z_1)(3+Z_1+2Z_2)]^{1/2}$ (Bardeen 1973, Novikov \& Thorne 1973),
where the upper and lower signs are for co-rotation and counter-rotation respectively.

As the inner region of an accretion disk is very influenced by the rotation of 
black hole, rotation should be incorporated in theoretical studies.
There are observational indications that black holes could be rotating rapidly and thus to 
study the inner properties of an accretion disk rotation should be incorporated correctly.
Iwasawa et al. (1996) have argued that the variable iron K emission 
line in MCG-6-30-15 arises from the inner part of an
accretion disk and it is strongly related to the spin of the black hole. 
It has been argued from other observational point of view (Karas \& Kraus 1996,
Iwasawa et al. 1996) that central black holes in galactic nuclei is likely to
be rapidly rotating. Also for the observation of gravitomagnetic precession, the inner edge 
of accretion disk is responsible (Markovic \& Lamb 1998, Stella \& Vietri 1998).
The temporal properties of the system are expected to depend on the inner edge of disk
which in turn depends on the rotation of the black hole.
The predictions of the disk properties will be incorrect if the pseudo-Newtonian 
modeling does not take into account spin of the black hole. 

The aim of this paper is to present a pseudo-Newtonian potential which can reproduce
exactly or in good agreement all the (inner) accretion disk properties close to the 
equatorial plane in Kerr geometry.
The potential should reproduce those features of a rotating black hole geometry
which have been reproduced by Paczy\'nski \& Wiita (1980)  potential (Eqn. \ref{pw}) 
for a non-rotating black hole. Thus, we will establish our potential in a same spirit as
Paczy\'nski and Wiita did for a non-rotating black hole. 
All other forms of the potential have been introduced without clear relation to the 
space-time metric. Here we will formulate our pseudo-potential
from the Kerr metric. As the metric is
involved directly to our calculation many of the features 
of Kerr geometry are inherent in our potential by design. In the next section we present the basic
equations and derive the pseudo-potential. In \S 3, we compare
a few results of the Kerr geometry with that of the potential and
in \S 4, we make our conclusions.

\section{Basic Equations and Pseudo-Potential}

The Lagrangian density for a particle in the Kerr space-time in Boyer-Lindquist coordinate
at the equatorial plane ($\theta=\pi/2$) can be written as
\begin{eqnarray}
\nonumber
2{\cal L}=-\left(1-\frac{2GM}{c^2r}\right){\dot t}^2-\frac{4GMa}{c^3r}{\dot t}{\dot \phi}
+\frac{r^2}{\Delta}{\dot r}^2\\+\left(r^2+\frac{a^2}{c^2}+
\frac{2GMa^2}{c^4 r}\right){\dot \phi}^2,
\label{l}
\end{eqnarray}
where over-dots denote the derivative with respect to the proper-time $\tau$ and
$\Delta=r^2+a^2/c^2-2GMr/c^2$.

The geodesic equations of motion are 
\begin{eqnarray}
E&=&{\rm constant}=\left(1-\frac{2GM}{c^2r}\right){\dot t}+\frac{2GMa}{c^3 r}
{\dot \phi},
\label{g1}
\end{eqnarray}
\begin{eqnarray}
\nonumber
\lambda &=&{\rm constant}=-\frac{2GMa}{cr}{\dot t}+
\left(r^2+\frac{a^2}{c^2}+\frac{2GMa^2}{c^4 r}\right){\dot \phi}.\\
\label{g2}
\end{eqnarray}
For the particle with non-zero rest mass $g_{\mu\nu}p^\mu p^\nu=-m^2$ (where $p^\mu$
is the momentum of the particles and $g_{\mu\nu}$ is the metric). Replacing
the solution for $\dot t$ and $\dot \phi$ from (\ref{g1}) and (\ref{g2}) into (\ref{l}) gives a differential 
equation for $r$ 
\begin{eqnarray}
\nonumber
\left(\frac{dr}{d\tau}\right)^2=\left(1+\frac{a^2}{c^2 r^2}+\frac{2GMa^2}{c^4 r^3}\right)E^2
-\left(1-\frac{2GM}{c^2 r}\right)\frac{\lambda^2}{r^2}\\-\frac{4GMaE\lambda}{c^3 r^3}
-\frac{m^2\Delta}{r^2}=\Psi.
\label{si}
\end{eqnarray}
Here, $\Psi$ can be identified as an effective potential for the radial geodesic motion.
The conditions for circular orbits are
\begin{eqnarray}
\Psi=0,\hskip1.cm\frac{d\Psi}{dr}=0.
\label{sie}
\end{eqnarray}
Solving for $E$ and $\lambda$ from (\ref{sie}) we get
\begin{eqnarray}
\frac{E}{m}=\frac{r^2-2GMr/c^2+a\sqrt{GMr/c^4}}{r(r^2-3GMr/c^2+2a\sqrt{GMr/c^4})^{1/2}},
\label{ee}
\end{eqnarray}
and
\begin{eqnarray}
\frac{\lambda}{m}=\frac{\sqrt{GMr/c^2}(r^2-2a\sqrt{GMr/c^4}+a^2/c^2)}{r(r^2-3GMr/c^2+2a\sqrt{GMr/c^4})^{1/2}}.
\label{le}
\end{eqnarray}
Equations (10) and (11) have been derived by Bardeen (1973).
Now as standard practice, we can define the Keplerian angular momentum distribution 
$\lambda_K=\frac{\lambda}{E}$. Therefore,
corresponding centrifugal force in Kerr geometry can be written as
\begin{eqnarray}
\frac{\lambda_K^2}{x^3}=\frac{(x^2-2a\sqrt{x}+a^2)^2}{x^3(\sqrt{x}(x-2)+a)^2}=F_x.
\label{sf}
\end{eqnarray}
Thus from above, $F_x$ can be identified as the gravitational force of black hole
at the Keplerian orbit.
The above expression reduces to Paczy\'nski-Wiita form for
$a=0$. Thus we propose Eqn. (\ref{sf}) is the most general form of the
gravitational force corresponding to the pseudo-potential in accretion disk 
around black holes. The general form of the corresponding pseudo-potential 
(which is $V_x=V_4=\int F_x dx$) is algebraically complicated, but simplifies for
any given value of $a$. In Appendix the
general form of the potential and its reduced form for a few particular Kerr
parameters are given.


\section{Comparison of the Results for Kerr geometry and Pseudo-Potential}

To establish the validity of this potential, we raise the following questions.
(1) Does this potential ($V_4$) reproduce the values of $x_b$ and $x_s$ as same as Kerr geometry? (2) Does it give the
correct value of $E_s$ as around of Kerr black hole? (3) How does the corresponding 
dissipation energy distribution $\eta(x)$ in the accretion disk by this potential 
match with that of pure general relativistic result? 

Apart from that, one can ask how simple the form of this potential so that it is applicable for 
other studies where an analytical form is required? Below, we are discussing all the questions 
one by one. If we can show,
our potential tackle all the above issues fairly well, we can conclude that
this is one of the best potential for an accretion disk around rotating back holes
as well as non-rotating ones.

At the marginally bound orbit mechanical energy $E$ reduces to zero and we get
\begin{eqnarray}
\frac{v^2}{2}+V=\frac{x}{2}\frac{dV}{dx}+V=0,
\label{rb}
\end{eqnarray}
which can be used to calculate $x_b$ for $V_4$. 
For the stability of an orbit $d\lambda/dx\ge0$, which for the specific potential 
$V_4$ is
\begin{eqnarray}
\nonumber
-3a^4+14a^3 \sqrt{x}+(x-6)x^3+6ax^{3/2}(x+2)\\-2a^2x(x+11)\ge0.
\label{rs}
\end{eqnarray}
The solution of Eqn. (\ref{rs}) with `equals to' sign gives the location 
of the last stable circular orbit ($x_s$)
for $V_4$. For any $a$, the $x_s$ computed from the above equation (\ref{rs}) 
matches exactly with the radius of last stable circular 
orbit in Kerr geometry. We are not reporting the $x_s$ for various $a$ values
as it is available in standard literatures. In Table-1 and 2 we list $x_b$
and $E_s$ for the potential $V_4$ and Kerr geometry for various values of $a$.

\vskip0.2cm
{\centerline{\large Table-1}}
{\centerline{\large Values of $x_b$}}
\begin{center}
{
\vbox{
\begin{tabular}{llllllllllllllllllllllllllllllll}
\hline
\hline
 $a$ & $0$ &$0.1$ &  $0.3$ & $0.5$ & $0.7$ & $0.998$   \\
\hline
\hline
 $V_4$  & $4.0$ &$3.788$ & $3.347$   & $2.870$  &  $2.333$ & 1.037 & \\
\hline
 Kerr & $4.0$ & $3.797$&  $3.373$ & $2.914$  & $2.395$ &  $1.091$  \\
\hline
\hline
 $a$ & $0$  & $-0.1$& $-0.3$ & $-0.5$ & $-0.7$ & $-0.998$  \\
\hline
\hline
 $V_4$  & $4.0$ & $4.206$ & $4.606$ &  $4.993$ &  $5.368$ &  $5.911$ \\
\hline
 Kerr & $4.0$ & $4.198$&  $4.580$  & $4.949$  & $5.308$ &  $5.825$ \\
\hline
\hline
\end{tabular}
}}
\end{center}
\begin{table*}[htbp]
{\centerline{\large Table-2}}
{\centerline{\large Values of $E_s$}}
\begin{center}
{
\vbox{
\begin{tabular}{lllllllllllllllllll}
\hline
\hline
 $a$ & $0$ &$0.1$ & $0.3$ & $0.5$ & $0.7$  & $0.998$   \\
\hline
\hline
 $V_4$  & $-0.0625$ &$-0.0663$ & $-0.0761$  & $-0.0904$ & $-0.1149$ & $-0.3533$   \\
\hline
 Kerr & $-0.0571$ & $-0.0606$&  $-0.0693$ & $-0.0821$ & $-0.1036$ & $-0.3209$  \\
\hline
\hline
 $a$ & $0$  & $-0.1$ & $-0.3$  &$-0.5$  &$-0.7$  &$-0.998$  \\
\hline
\hline
 $V_4$  & $-0.0625$ & $-0.0591$ & $-0.0536$ & $-0.0491$ & $-0.0454$ & $-0.0409$ \\
\hline
 Kerr & $-0.0571$ & $-0.0542$  & $-0.0492$  &$-0.0451$  &$-0.0418$ & $-0.0378$ \\
\hline
\hline
\end{tabular}
}}
\end{center}
\end{table*}

From Table-1, it is clear that for all values of $a$, $V_4$ can reproduce the value of $x_b$ in very good
agreement with general relativistic results. The maximum error in $x_b$ is $\sim5\%$.
Table-2 indicates that $V_4$ produces $E_s$ in a fairly good agreement with Kerr geometry
with a maximum possible error $\sim 10\%$. Thus, the potential $V_4$ will product a slightly
larger luminosity than the general relativistic one in the accretion disk for 
a particular accretion rate.  Note that, for counter-rotating black holes the errors are less 
than those of a co-rotating one. 

Next we will compare the total energy dissipation $\eta$ in the accretion disk
for this potential with that of general relativistic result. We choose a simple $\alpha$-disk model to compute
pseudo-Newtonian $\eta(x)$ (Bj\"ornsson \& Svensson 1991, Frank et al. 1992, Artemova 1996), 
where for different choices of pseudo-potential, $\Omega$ can be 
different which could have been reflected to the final profile of $\eta(x)$ (Bj\"ornsson \& Svensson 
1991, 1992). Following Novikov \& Thorne (1973), Page \& Thorne (1974) and Bj\"ornsson (1995)
we calculate the corresponding general relativistic $\eta$ profile. 
In Fig. 1, we show $\eta/{\dot m}$ ($\dot m$ is the accretion rate) 
as a function of $x$, for different values of Kerr parameter $a$. Here also our pseudo-Newtonian results agree within 
$10\%$ of general relativistic values. For moderate rotation almost there is no error,
while for very rapidly rotating black holes the deviation increases but still within $10\%$.

Apart from that, the analytical expression for the force (Eqn. (\ref{sf})) as well as 
potential at given values of $a$ (Eqns. (\ref{v1}) and (\ref{vp5})) are relatively
simple which will make it easy to implement in other applications like detailed fluid dynamical studies, 
analysis of the parameter space in disk etc. Thus this potential
satisfies all the criteria for a good pseudo-potential which can describe an accretion
disk using non-relativistic equations.


\begin{figure*}[htbp]
\centerline{\psfig{file=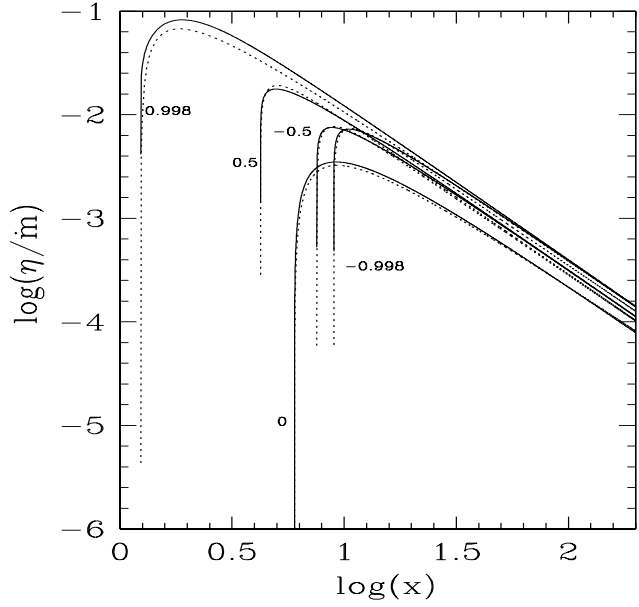,width=14cm,height=14.0cm}}
\figcaption{Energy dissipation per unit accretion rate $\left(\eta/{\dot m}\right)$ as a function
of radial coordinate ($x$) for various values of specific angular momentum of the
black hole ($a$), which are indicated at each set. For each set, solid and dotted curves
show results of general relativity and our potential respectively.}
\label{fig1}
\end{figure*}


\section{Conclusions}

We have prescribed a general pseudo-potential for the modeling of accretion disks around
rotating black holes. For the non-rotating case, it reduces to Paczy\'nski-Wiita
potential. Unlike previous works, this potential is derived from the metric (Kerr geometry)
at the equatorial plane. Naturally it exhibits better accuracy as it is derived from the metric itself.
Following the same procedure, using Schwarzschild metric one can derive
Paczy\'nski-Wiita potential. The detailed calculations of various 
geodesic equation are available in standard literatures (e.g. Shapiro \& Teukolsky 1983). 
Our potential is valid for both co-rotating and counter-rotating black holes which is not
necessarily the case for earlier potentials (Artemova et al. 1996).
In fact for our pseudo-Newtonian potential, the counter-rotating results agree better with the general 
relativistic ones, presumably because of the larger values of $x_b$ and $x_s$ with respect to
that of co-rotating cases.  But, still for our potential the possible
error is $10\%$ at most for any rotation. Thus the inference of various observational aspects using our 
potential may have better accuracy. If the description of disk property is 
acceptable within $10\%$ accuracy, our potential should be recommended. 
It should be mentioned that, although this pseudo-Newtonian potential is applicable close to the equatorial plane, 
it may not be a good approximation to follow light rays, orbits far from the equator. As because 
at the very beginning of our calculation we have chosen $\theta=\pi/2$ this further constraint arises. 
Such a pseudo-Newtonian potential for generalized $\theta$ has not been proposed before and hence
it might be useful to derive that one following the method prescribed in this work.

Next, one can apply this potential for various fluid dynamical problems.
It is shown here that the maximum error in calculation of $\eta$ 
is $10\%$ even for the rapidly rotating black holes. One should study:
how does the rotation of black hole affect the fluid properties in an accretion disk? 
How does it affect the parameter region of disk? In a next work, we expect to explore all these issues.

\appendix
\section{Appendix: Analytical Expressions of Pseudo-Potential }

As we have an analytical form of the gravitational force (\ref{sf}), we can calculate
the corresponding potential as $V_x=\int F_x dx$. Thus the most general 
expression for the pseudo-potential is

\begin{eqnarray}
\nonumber
V_x&=&-\frac{a^2}{2x^2}+\frac{4a}{\sqrt{x}}+\frac{2(9a^3 x-10ax+16\sqrt{x}-13a^2\sqrt{x}+6a^3
-8a)}{(27a^2-32)(x^{3/2}-2\sqrt{x}+a)}-2log(x)\\
\nonumber
&+&\frac{2}{27a^2-32}\sum_{y=x_1,x_2,x_3}\left[\frac{1}{3y^2-2}
log(\sqrt{x}-y)(54a^2 y^2-64y^2+63a^3 y\right.\\
&-&\left. 74ay-107a^2+128)\right],
\label{gv}
\end{eqnarray}
where,
\begin{eqnarray}
\nonumber
x_1=\frac{2^{4/3}}{p}+\frac{p}{2^{1/3} 3},\hskip0.3cm x_2=-\left(\frac{2^{1/3}q}{p}+
\frac{pq^*}{2^{1/3}6}\right),\hskip0.3cm x_3=x_2^*,\\
{\rm and}\hskip0.3cm p=(\sqrt{729a^2-864}-27a)^{1/3},\hskip0.3cm q=(1+i\sqrt{3}).
\label{xroot}
\end{eqnarray}
Here, '$*$' denotes the complex conjugate.
Though Eqn. (\ref{gv}) looks very complicated if we specify the particular values of
$a$ it reduces to rather simpler expression. For example, if we choose $a=0$, it
reduces to Paczy\'nski-Wiita potential ($-\frac{1}{x-2}$). For some other values of
$a$ the analytical forms of the potential reduce as
\begin{eqnarray}
V_x^{a=\pm 1}=-\frac{1}{2x^2}\pm\frac{4}{\sqrt{x}}-\frac{2}{5}
\left[\frac{(\mp x+3\sqrt{x}\mp 2)}{(x^{3/2} -2\sqrt{x}\pm 1)}+log\left(\frac{(\sqrt{x}\pm 
B)^A}{(\sqrt{x} \mp D)^C}\right)\right]-2log(x),
\label{v1}
\end{eqnarray}
where, $A=2.15542$, $B=1.61803$, $C=12.1554$, $D=0.618034$
and
\begin{eqnarray}
\nonumber
V_x^{a=\pm 0.5}&=&-\frac{1}{8x^2}\pm\frac{2}{\sqrt{x}}-E\left[\frac{\mp 7.75x+25.5\sqrt{x}
\mp 6.5}{2x^{3/2}-4\sqrt{x}\pm 1}\right.\\ &+& \left. log\left(\frac{(\sqrt{x}\pm G)^F}{(\sqrt{x}\mp I)^H
(\sqrt{x}\mp K)^J} \right)\right]-2log(x),
\label{vp5}
\end{eqnarray}
where, $E=0.0792079$, $F=5.64616$, $G=1.52569$, $H=5.93863$, $I=1.26704$, $J=50.2075$,
$K=0.258652$. As $a\rightarrow 1$, $H\rightarrow 0$. One can easily check that both 
(\ref{v1}) and (\ref{vp5}) asymptotically vary as $-1/x$. The expressions (\ref{v1}) 
and (\ref{vp5}) could have been reduced to a more simpler form if we approximate
the values of decimal number in constants. But in this manner accuracy of the solution
would be reduced.

\begin{acknowledgements}
I would like to thank the anonymous referee for various suggestions which improved the presentation
of the paper. I also would like to thank Ranjeev Misra for discussions.
\end{acknowledgements}



{}

\end{document}